\documentclass[]{aastex631}

\newcommand{\citefull}[1]{\citeauthor{#1} \citeyear{#1}}
\usepackage{bm}
\usepackage{amsmath}
\let\tablenum\relax
\usepackage{siunitx}
\usepackage{graphicx}

\begin{document}

\title{Predicting the detection yields of giant planets and brown dwarfs with CSST astrometry}

\author[0009-0000-9341-3442]{Yifan Xuan}
\altaffiliation{yifan.xuan@sjtu.edu.cn}
\affiliation{Tsung-Dao Lee Institute, Shanghai Jiao Tong University, 1 Lisuo Road, Shanghai 201210, People’s Republic of China}

\author[0000-0001-6039-0555]{Fabo Feng}
\altaffiliation{ffeng@sjtu.edu.cn}
\affiliation{Tsung-Dao Lee Institute, Shanghai Jiao Tong University, 1 Lisuo Road, Shanghai 201210, People’s Republic of China}
\affiliation{School of Physics and Astronomy, Shanghai Jiao Tong University, 800 Dongchuan Road, Shanghai 200240, People’s Republic of China}

\author[0009-0000-1630-3725]{Zhensen Fu}
\affiliation{Shanghai Astronomical Observatory, Chinese Academy of Sciences, Shanghai 200030, China}
\affiliation{School of Astronomy and Space Sciences, University of Chinese Academy of Sciences (UCAS), 19A Yuquan Road, Beijing 100049, China}

\author[0000-0002-9346-0211]{Shilong Liao}
\affiliation{Shanghai Astronomical Observatory, Chinese Academy of Sciences, Shanghai 200030, China}
\affiliation{School of Astronomy and Space Sciences, University of Chinese Academy of Sciences (UCAS), 19A Yuquan Road, Beijing 100049, China}

\author{Zhaoxiang Qi}
\affiliation{Shanghai Astronomical Observatory, Chinese Academy of Sciences, Shanghai 200030, China}
\affiliation{School of Astronomy and Space Sciences, University of Chinese Academy of Sciences (UCAS), 19A Yuquan Road, Beijing 100049, China}

\author[0000-0002-3759-1487]{Yang Chen}
\affiliation{School of Physics and optoelectronic engineering, Anhui University, Hefei 230601, China}
\affiliation{National Astronomical Observatories, Chinese Academy of Sciences, 20A Datun Road, Beijing 100101, China}




\begin{abstract}
Chinese Space Station Telescope (CSST), which will begin its scientific operations around 2027, is going to survey the sky area of the median-to-high Galactic latitude and median-to-high ecliptic latitude. 
The high astrometric precision of the CSST Survey Camera for faint objects enables the detection of a number of giant planets and brown dwarfs around M-dwarfs and brown dwarfs via differential astrometry in its optical survey. 
In this paper, we predict the number of giant planets and brown dwarfs around stars and brown dwarfs detectable with CSST astrometry. 
We generate synthetic samples of CSST stellar and substellar sources, and carry out companion injection-recovery simulations in the samples using different occurrence rates for FGK-dwarfs, M-dwarfs and brown dwarfs. 
We calculate companion yields based on CSST astrometric precision. 
Our analysis reveals that over its 10-year mission, the CSST Survey Camera could barely discover giant planets and low-mass BDs around FGK-dwarfs, but is projected to detect $20 - 170$ giant planets and low-mass brown dwarfs around M-dwarfs within \SI{300}{pc}, and $300 - 570$ brown dwarf binaries within \SI{600}{pc}.
Therefore, CSST astrometry is likely to significantly increase the current sample of substellar companions around M-dwarfs and brown dwarfs. 
This sample will deepen our understanding of planet formation and evolution around low-mass stars and brown dwarfs. 
\end{abstract}

\keywords{Exoplanet astronomy(486) --- Low mass stars(2050) --- Brown dwarfs(185) --- Astrometric exoplanet detection(2130) --- Space Astrometry(1541) --- Space telescopes(1547)}


\section{Introduction} \label{sec:intro}


The search for long-period massive companions around M-dwarfs and brown dwarfs is of great scientific interest in recent years.
M-dwarfs are the smallest, dimmest, and most abundant type of stars in the Milky Way. 
Substellar companions around M-dwarfs, which encompass both brown dwarfs and giant planets, are of particular interest, and searching for giant exoplanets around M-dwarfs has gained increasing attention recently \citep{Kanodia2024}. 
Previous simulations based on bottom-up planet formation mechanisms such as pebble accretion suggest that less massive stars tend to host smaller planets \citep{Liu2020}. 
However, giant planets and brown dwarfs have been detected through various methods around low-mass stars such as GJ3512 b \citep{Morales2019}, TOI-5205 b \citep{Kanodia2023}, GJ463 b \citep{Endl2022, Sozzetti2023} and Gaia-5 b \citep{Stefansson2025}. 
Those observations indicate alternative scenarios such as disk instability \citep{Boss2023} or pebble accretion plus planet-planet collisions \citep{Pan2024}. 
The key to distinguishing between various formation scenarios is to detect a statistically significant sample of giant planets and brown dwarfs around M-dwarfs for robust population synthesis. 

In the past few years, more attention has been paid to substellar companions around brown dwarfs (BDs) whose masses are even smaller than M-dwarfs. 
Brown dwarf binaries are found to be less prevalent than stellar binaries, but they have higher mass ratios and more tightly bound orbits than their stellar analogues \citep{Gizis2003, Burgasser2006, Fontanive2018, DeFurio2022, Factor2023, Fontanive2023}. 
The differences of binary fractions, mass ratios and separations between brown dwarfs and stars indicate different formation mechanisms. 
The formation mechanism for brown dwarf binaries \textemdash \ whether turbulent fragmentation \citep{Padoan2004}, disk instability \citep{Goodwin2007}, ejection from stellar systems \citep{Reipurth2001} or circumstellar disk encounters \citep{Shen2010,Thies2010, Fu2025} \textemdash \ remains uncertain.
Furthermore, masses of isolated brown dwarfs are inferred from evolutionary models while masses of those in binaries could be estimated from dynamical measurements. 
Dynamical masses of brown dwarfs could help to test substellar evolutionary models and atmospheric assumptions \citep{Brandt2021, Xuan2024}.

\begin{figure*}[!ht]
\centering
\includegraphics[width=0.75 \textwidth]{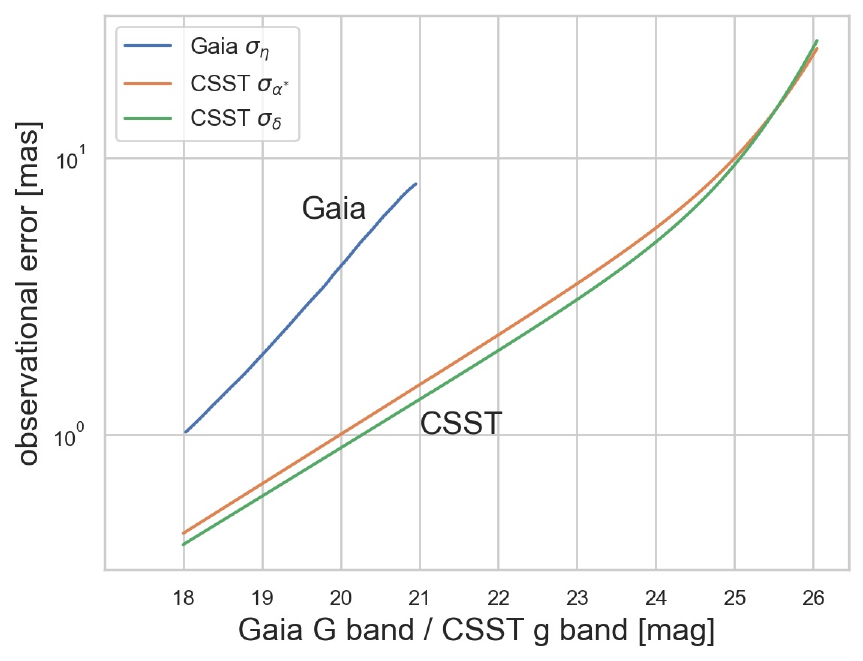}
\caption{Comparison of the astrometric precision of one single measurement by Gaia and CSST. The astrometric precision in the R.A. and Decl. directions are shown with different colors. The horizontal axis is either the Gaia G band or the CSST g band. Gaia position uncertainties are derived from equation (1-4) in \cite{Perryman2014} and equation (7) in \cite{Gaia2016} while CSST position uncertainties are obtained from figure 4 in \cite{Fu2023}.}
\label{fig:comparison}
\end{figure*}

Nevertheless, the current sample size is inadequate to test different formation scenarios for giant planets around M-dwarfs and brown dwarf binaries. 
Although terrestrial planets are ubiquitous around low-mass stars \citep{Dressing2013, Dressing2015, Ment2023}, long-period giant planets are rarely detected around M-dwarfs compared with those around solar-like stars \citep{Sabotta2021, Pass2023, Mignon2025}.
At present, most of the substellar companions around M-dwarfs are detected by microlensing (e.g., \citefull{Zang2021}), transit, and radial velocity (e.g., \citefull{Canas2022}, \citefull{Carmichael2022}). 
However, microlensing is sensitive to wide-separation companions but can only provide limited orbital information \citep{Gaudi2012}, and the lens systems in microlensing events are typically rather faint and far away from Earth, making it hard for follow-up observations and detailed characterizations. 
Transit is generally suitable for edge-on companions in close-in orbit. 
Brown dwarf binaries are usually detected by imaging \citep{Gizis2003, Close2003, Reid2006, Burgasser2006, Stumpf2010, Aberasturi2014, Fontanive2018, DeFurio2022, Fontanive2023} and radial velocity \citep{Basri2006, Joergens2008, Blake2010}. 
But imaging can only get access to model-dependent masses, and radial velocity is only able to measure minimum companion masses.

As a supplement to the techniques above, the astrometry method can detect long-period substellar companions around stars and brown dwarfs, and determine all orbital parameters if the position measurement is precise enough and the observing timespan is long enough. 
As shown in \cite{Feng2024}, the astrometric signature of a reflex motion induced by a companion is proportional to the companion mass and semi-major axis, and inversely proportional to the host mass and system distance. 
Therefore, an astrometric survey of nearby M-dwarfs may complement microlensing, transit, radial velocity and imaging techniques, and unveil the lurking substellar companions in the solar neighborhood. 
Gaia is a cornerstone mission of space astrometry, which can measure the positions of stars with standard errors of a few dozen microarcseconds at the end of its mission \citep{Gaia2016, Gaia2023b}. \cite{Sozzetti2014} anticipated that Gaia could detect 100 giant planets around nearby M-dwarfs, and \cite{Perryman2014} predicted that Gaia could discover 1,000 
\textendash 1,500 massive long-period planets around M-dwarfs within \SI{100}{pc}.

Complementary to Gaia's absolute astrometry for all-sky stars brighter than $G < $ \SI{21}{mag}, CSST will measure differential astrometry for objects as faint as \SI{26}{mag}. The \SI{2}{m}-aperture Chinese Space Station Telescope (CSST) is a major science project of China Manned Space Program. 
CSST is expected to start science operations around 2027 and has a nominal mission duration of 10 years \citep{Zhan2011, Zhan2018, Zhan2021, CSST2025}. 
During its observations, it will fly independently in the same orbit as the space station, which is about \SI{400}{km} in height, to dock for refueling and servicing while maintaining a large distance apart \citep{Su2014}. 
The scientific goals of CSST include investigating a range of significant questions in cosmology, galaxies, stars, and exoplanets. To achieve these goals, CSST will be equipped with five instruments including the Survey Camera, Terahertz Receiver, Multichannel Imager, Integral Field Spectrograph and Cool-Planet Imaging Coronagraph \citep{Zhan2021, CSST2025}. 
Although precision cosmology is the main science driver of the CSST optical survey \citep{Gong2019}, the design of the CSST Survey Camera (CSST-SC) is suitable for astrometric studies of objects fainter than 20 magnitudes \citep{Fu2023}, which provides a great opportunity to detect long-period massive substellar companions around nearby faint objects like M-dwarfs and brown dwarfs.
The comparison of astrometric uncertainties for one single measurement between Gaia and CSST is shown in Figure \ref{fig:comparison}. 
Although the final astrometric precision of CSST is not better than that of Gaia for objects brighter than 21 mag due to a smaller number of observations, CSST's fainter magnitude range and larger field of view enable it to complement the Gaia stellar catalogue and detect substellar companions around faint objects beyond Gaia's limiting magnitude.

This paper is structured as follows. 
In Section \ref{sec:sample}, we describe the CSST survey strategy and the establishment of the synthetic CSST stellar and substellar samples. 
Section \ref{sec:model} introduces the adopted occurrence rate models and describes the companion injection-recovery simulations in the framework of astrometry based on CSST's astrometric precision. 
We present results of the simulations in Section \ref{sec:results} and finally conclude in Section \ref{sec:conclusions}. 

\begin{figure*}[!ht]
\centering
\includegraphics[width=0.75 \textwidth]{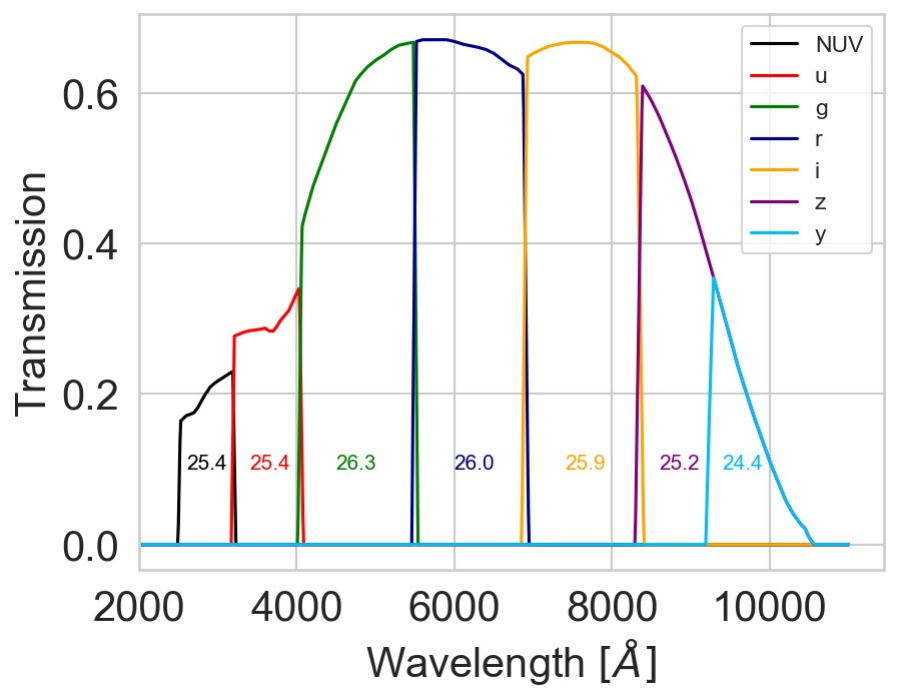}
\caption{CSST filter transmission curves corresponding to NUV, u, g, r, i, z and y band. The limiting magnitude of the wide field survey for each band is 25.4, 25.4, 26.3, 26.0, 25.9, 25.2, 24.4 mag, shown under corresponding curves. This figure is adapted from figure 1 in \cite{Lu2024}.}
\label{fig:throughput_of_csst}
\end{figure*}

\section{Synthetic samples of CSST sources} \label{sec:sample}
\begin{figure*}[!ht]
\centering
\includegraphics[width=1.0 \textwidth]{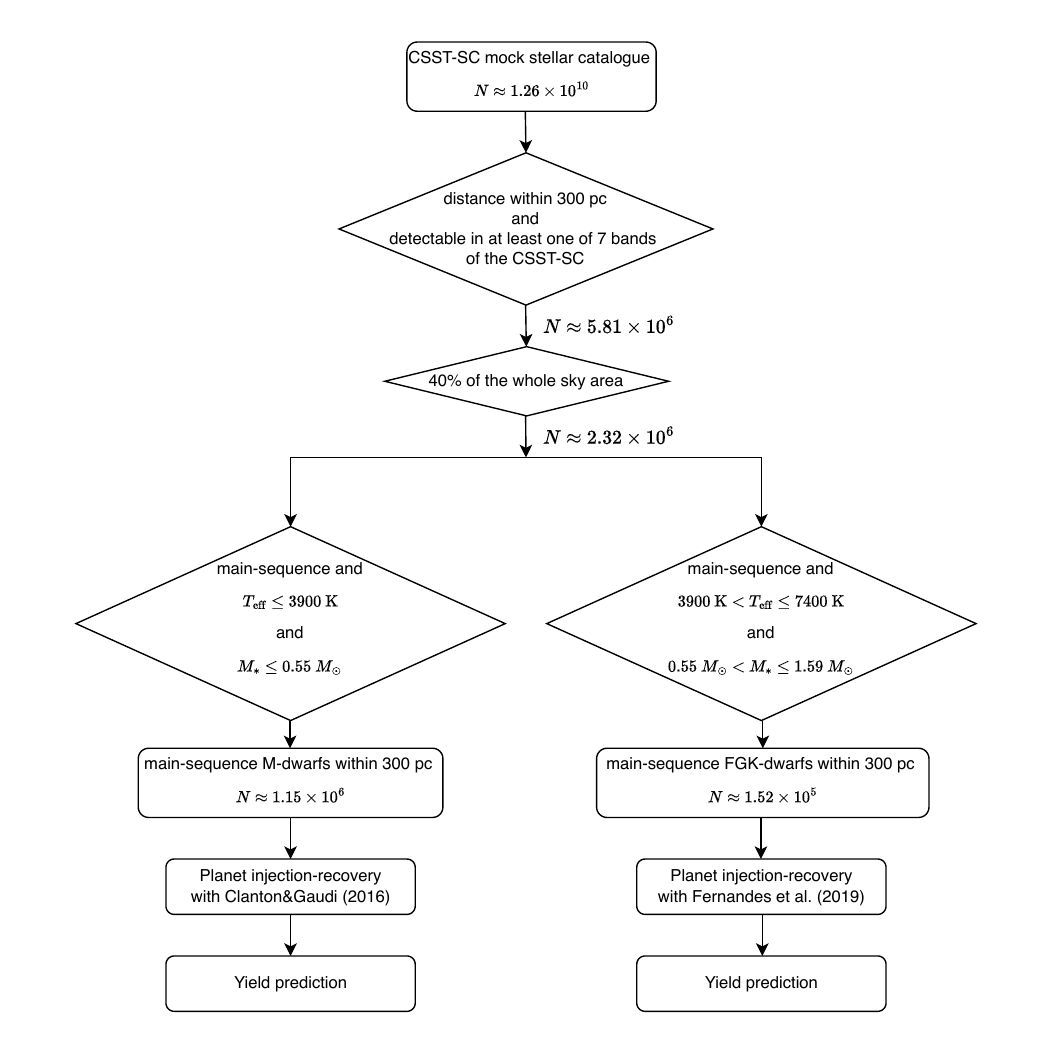}
\caption{Workflow for stars from constructing stellar synthetic samples of M-dwarfs and FGK-dwarfs to yield predictions.}
\label{fig:workflow_stars}
\end{figure*}

\begin{figure*}[!ht]
\centering
\includegraphics[width=1.0 \textwidth]{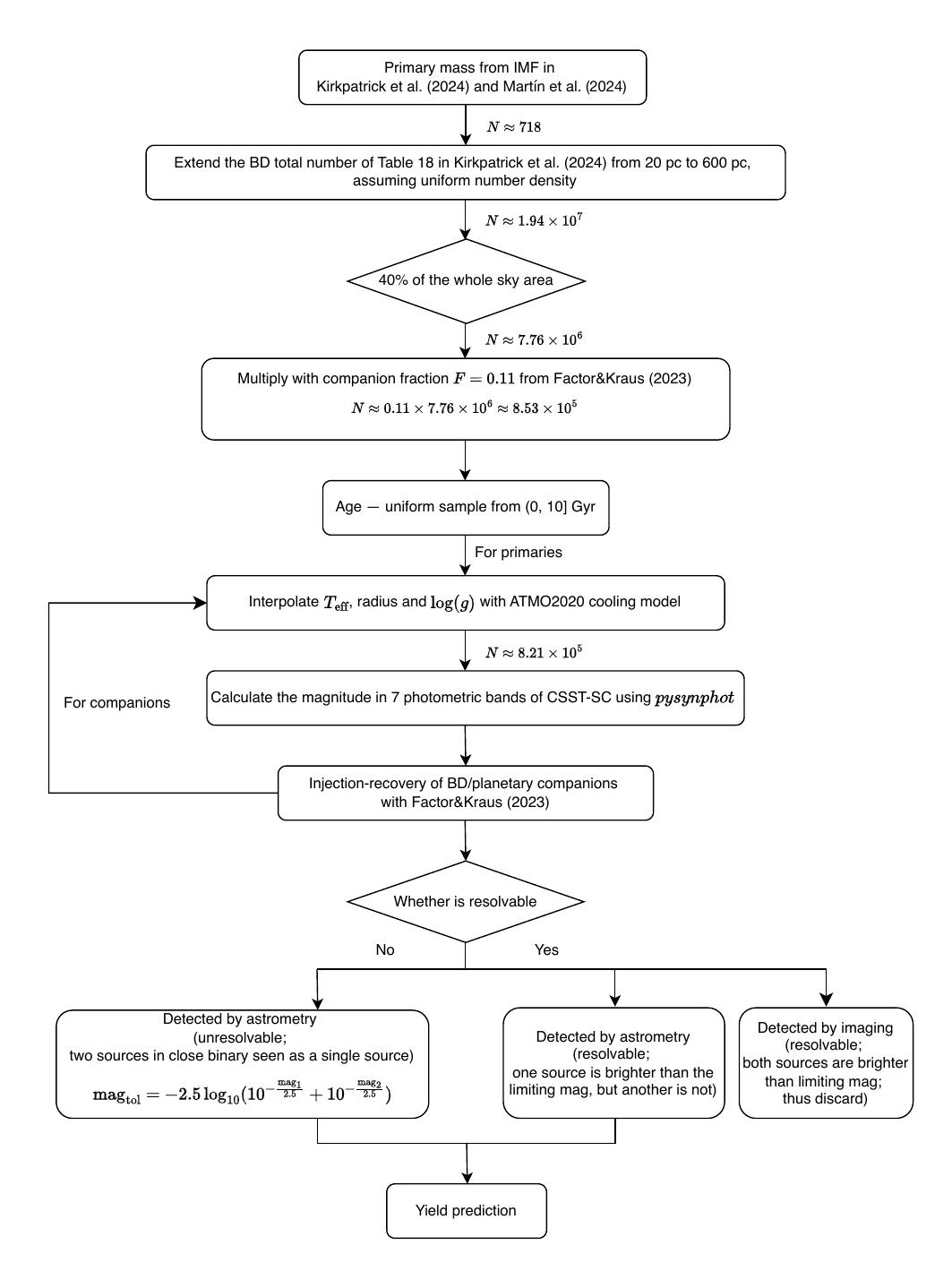}
\caption{Workflow for brown dwarfs from constructing substellar synthetic samples to yield predictions.}
\label{fig:workflow_BDs}
\end{figure*}

In this section, we introduce basic information of the CSST Survey Camera, its survey strategy and observing cadence in Section \ref{survey_strategy}. 
The processes of constructing the synthetic samples of CSST stellar and substellar sources are elaborated in Section \ref{stellar_sources} and \ref{substellar_sources}. 
The workflows of predicting the yield of substellar companions are shown in Figure \ref{fig:workflow_stars} and \ref{fig:workflow_BDs} for stars and brown dwarfs respectively, from the start of constructing synthetic samples of CSST sources to the end of yield predictions. 

\subsection{CSST survey strategy} \label{survey_strategy}
CSST will allocate ${\sim}70\%$ of its 10-year operation time for the Survey Camera to image approximately \SI{17500}{deg^{2}} of the sky in the median-to-high Galactic latitude \citep{Zhan2021, Fu2023}. 
The Survey Camera has seven filters ranging from \SI{255}{nm} to \SI{1100}{nm}, which include the NUV, u, g, r, i, z, and y bands covering a wavelength range from near-UV to near-IR. 
There are a total of 18 detectors on the Survey Camera for multiband photometry, including 4 for the NUV band, 4 for the y band, and 2 for each of the ugrizy bands, shown as figure 2 in \cite{CSST2025}. 
In CSST's 10-year mission duration, the \SI{17500}{deg^{2}} main sky survey areas will be observed once by each of the 18 detectors, while the \SI{400}{deg^{2}} selected deep fields will be observed four times by each detector. 
The exposure times for the main sky survey areas and the selected deep fields are \SI{150}{s} and \SI{250}{s}, respectively \citep{Zhan2021, CSST2025}. 
The transmission curves and limiting magnitudes of seven bands are shown in Figure \ref{fig:throughput_of_csst}. 


\subsection{Synthetic sample of CSST stellar sources} \label{stellar_sources}
\verb|TRILEGAL| is a commonly-used galactic stellar population synthesis model based on 2MASS and SDSS data \citep{Girardi2005, Girardi2012}, and \cite{Chen2023} has built a Milky Way mock stellar catalogue for the CSST-SC photometric system using \verb|TRILEGAL|.
The catalogue provided by \cite{Chen2023} is specifically designed for the CSST photometric system and encompasses the main stellar components of the Galaxy, including the thin disk, thick disk, halo, and bulge. 
It consists of approximately 12.6 billion stars, reaching magnitude down to g = \SI{27.5}{mag} in the AB magnitude system \citep{Chen2023}.

From the mock stellar catalog, we first select a sample of stars within \SI{300}{pc} that are detectable in at least one of the seven photometric bands of the CSST-SC. 
\cite{Perryman2014} and \cite{Sozzetti2014} set \SI{100}{pc} as the distance limit for Gaia, but here we expand this range to \SI{300}{pc} because of CSST's superior sensitivity to faint sources.  
Here ``detectable'' means the star should be brighter than the limiting magnitude and fainter than the limiting magnitude minus 8.5 in a given band to avoid saturation (see table 3 in \citefull{Zhan2021}). 

Then we randomly select 40\% of the stellar sample because the CSST-SC will observe 17500\,deg$^2$ of all sky \citep{Zhan2021, CSST2025}. 
After that, main-sequence (\verb|label=1|) M-type and FGK-type stars are selected. 
Pre-main-sequence, evolved, and early-type stars are excluded because their number is much smaller than main-sequence stars and their planet occurrence rates are poorly known compared with main-sequence late-type stars. 
FGK-dwarfs are filtered with the effective temperature $\SI{3900}{K} \textless T_{\mathrm{eff}} \leq \SI{7400}{K}$ and the stellar mass $0.55 \textless M_{*} \leq \SI{1.59}{M_{\odot}}$ while M-dwarfs are filtered with $T_{\mathrm{eff}} \leq \SI{3900}{K}$ and $M_{*} \leq \SI{0.55}{M_{\odot}}$. 
The property distributions of the synthetic samples of M-dwarfs are presented as grey lines in Figure \ref{fig:M-dwarfs_yield}. 
The number of FGK-dwarfs and M-dwarfs is $1.52 \times 10^{5}$ and $1.15 \times 10^{6}$, respectively, which are consistent with the number of stars within \SI{300}{pc} in Gaia DR3 ($2.87 \times 10^{5}$ and $1.18 \times 10^{6}$ respectively) \citep{Gaia2016, Gaia2023b}, so we don't scale the total number of the sample.

\subsection{Synthetic sample of CSST substellar sources}
\label{substellar_sources}
We also investigate the potential of CSST-SC to detect companions around substellar objects by astrometry. 
Since there is no ready-made mock catalogue of substellar objects for CSST, we establish a catalogue from the initial mass function in \cite{Kirkpatrick2024} and \cite{Martin2024}. 
\cite{Kirkpatrick2024} gives the initial mass function of a four-piece power law based on the full-sky \SI{20}{pc} census of $\sim 3600$ stars and brown dwarfs, but the minimum mass only reaches $\SI{50}{M_{\mathrm{Jup}}}$, an intermediate mass of the brown dwarf regime. 
Therefore, we also adopt the initial mass function from the latest Euclid observations of free floating new-born planets in the $\sigma$ Orionis cluster to go down to the planetary regime (figure 10 in \cite{Martin2024}). 
The combined initial mass function is given below
\begin{equation}
\xi(M) = \left\{
\begin{aligned}
& C_{1} M^{-\alpha_{1}}, \ 0.05 \textless M \leq 0.075 M_{\odot} \\
& C_{2} M^{-\alpha_{2}}, \ 0.011 \textless M \leq 0.05 M_{\odot} \\
& C_{3} M^{-\alpha_{3}}, \ 0.003 \textless M \leq 0.011 M_{\odot}
\end{aligned}
\right.
\end{equation}
where $C_{1}=0.134, \ \alpha_{1}=0.25, \ C_{2}=0.165 , \ \alpha_{2}=0.18, \ C_{3}=0.215, \ \alpha_{3}=0.12$.

Assuming uniform number density, we extend the distance of the total number of objects with mass less than $\SI{75}{M_{\mathrm{Jup}}}$ in table 18 of \cite{Kirkpatrick2024} from \SI{20}{pc} to \SI{600}{pc}, and then randomly select 40\% of the sample considering CSST's survey areas. 
The companion fraction of brown dwarfs is $10\% \sim 20\%$ according to previous studies \citep{Gizis2003, Close2003, Reid2006, Basri2006, Joergens2008, Aberasturi2014, Fontanive2018, DeFurio2022, Factor2023}, and we adopt the companion fraction of $F=0.11^{+0.04}_{-0.03}$ in \cite{Factor2023}. The ages of substellar objects in the catalogue are uniformly sampled from $0 - \SI{10}{Gyr}$. 
Given the age and mass, we derive $T_{\mathrm{eff}}, \ \log(g)$ and the radius of every substellar object in the catalogue by interpolating ATMO 2020 evolutionary models \citep{Phillips2020} assuming chemical equilibrium.
With the parameters derived above and the transmission curves of seven photometric bands on the CSST-SC \citep{Zhan2021}, we calculate the apparent magnitudes of every substellar object in the NUV, u, g, r, i, z and y bands using \verb|PySynphot|, which is a synthetic photometry package developed for HST \citep{pysynphot2013}. 

The total number of substellar objects in the synthetic sample is $8.21 \times 10^{5}$, and the property distributions of CSST substellar sources are presented as grey lines in Figure \ref{fig:BDs_yield}.
Most of brown dwarfs and giant planets are faint in the ultraviolet and optical band but brighter in infrared. 

\section{Model} \label{sec:model}

\subsection{Companion occurrence rate model}
\subsubsection{For FGK-dwarf hosts}
We adopt three different occurrence rate models for long-period, massive companions around FGK-dwarfs, M-dwarfs and brown dwarfs. 
\cite{Fernandes2019} computed occurrence rates of giant planets around FGK-dwarfs based on the planet detections and simulated survey completeness from the HARPS and CORALIE radial velocity surveys \citep{Mayor2011}.
They derived a broken power-law distribution of giant planet occurrence rates as a function of orbital period and planet mass expressed as
\begin{equation} \label{eq: GP occurrence rate around FGK-dwarfs}
f(M,P) = \frac{\mathrm{d}^{2}N}{\mathrm{d} \log P \mathrm{d} \log M} = c_{0} (\frac{M}{10 M_{\oplus}})^{m_{1}} 
\left\{
\begin{aligned}
& (\frac{P}{P_{\mathrm{break}}})^{p_{1}}, \ P \leq P_{\mathrm{break}} \\
& (\frac{P}{P_{\mathrm{break}}})^{p_{2}}, \ P \textgreater P_{\mathrm{break}}
\end{aligned}
\right .
\end{equation}
where $c_{0}=0.84^{+0.18}_{-0.15}$ is a normalization factor, $m_{1}=-0.45 \pm 0.05$ is the power-law index of the planet mass distribution, $P_{\mathrm{break}} = 1581^{+894}_{-392}$ $\mathrm{d}$ is the location of the break in the period distribution, and $p_{1}=0.65^{+0.20}_{-0.15}$ and $p_{2}=-0.65^{+0.20}_{-0.15}$ are the power-law indices before and after the break respectively. 
The ranges of planet mass and orbital period are $0.1 - \SI{20}{M_{\mathrm{Jup}}}$ and $10-\SI{10000}{d}$ respectively. 

\subsubsection{For M-dwarf hosts}
\cite{Clanton2016} synthesized results from five different exoplanet surveys using three independent detection methods including microlensing \citep{Gould2010, Sumi2011}, radial velocity \citep{Montet2014}, and direct imaging \citep{Lafrenière2007, Bowler2015}, and constructed a power-law form for the frequency of long-period planets around M-dwarfs. 
We adopt the non-parametric occurrence rate distribution from table 3 in \cite{Clanton2016} with planet mass ranging from $1$ to $\SI{10000}{M_{\oplus}}$ and orbital period ranging from $1$ to $\SI{100000}{d}$. 
We inject simulated companions for every star in the synthetic sample following a log-uniform distribution over the grids of the $P-m$ (orbital period versus planet mass) parameter space. 
For each node in the grid, we generate 100 sets of other orbital parameters, including eccentricity $e$, cosine of inclination $\cos i$, longitude of the ascending node $\Omega$, argument of pericentre $\omega$, and mean anomaly at epoch $M_{0}$, from uniform distributions. 

\subsubsection{For brown dwarf hosts}
\cite{Factor2023} derived log-normal separation and power-law mass-ratio distributions of the brown dwarf companion frequency based on archival HST/NICMOS\footnote{NICMOS stands for the Near Infrared Camera and Multi-Object Spectrometer on the Hubble Space Telescope.} surveys of field brown dwarfs. 
The expected brown dwarf companion frequency $R$ in a bin is expressed as
\begin{equation}  \label{eq: BD companion freq}
R(\log(\rho), q | F, \gamma, \overline{\log(\rho)}, \sigma_{\log(\rho)}) \Delta \log(\rho) \Delta q = \frac{\gamma+1}{\sqrt{2 \pi}\sigma_{\log(\rho)}} F q^{\gamma} \exp(-\frac{(\log(\rho) - \overline{\log(\rho)})^{2}}{2 \sigma^{2}_{\log(\rho)}}) \Delta \log(\rho) \Delta q
\end{equation}
where $\rho$ is the observed projected separation and $q$ is the mass ratio between the secondary and primary brown dwarfs. 
The binary population is characterized by a companion frequency $F=0.11^{+0.04}_{-0.03}$, a power-law mass ratio distribution with exponent $\gamma$, a log-normal projected separation distribution with mean $\overline{\log(\rho)}=0.34$ and standard deviation $\sigma_{\log(\rho)}=0.58$. 
It should be noted that the only parameter affected by the assumed brown dwarf age among the four is $\gamma$. 
Figure 7 in \cite{Factor2023} gives the value of $\gamma$ as a function of the assumed age between $0.9-\SI{3.1}{Gyr}$, and we linearly extrapolate the age range to $0-\SI{10}{Gyr}$. 
The ranges of mass ratio $q$ and projected separation $\rho$ are $0-1$ and $0.1-\SI{100}{au}$ respectively. 
Given the parameters above, the apparent magnitudes of companions in seven bands are also calculated using \verb|PySynphot| like primaries.

\subsection{Astrometric model}
CSST-SC uses differential astrometry to measure positions of target stars with respect to reference stars \citep{Cameron2009, Muterspaugh2010}, whose precise positions are provided by Gaia absolute astrometry.
Gaia provides much more precise astrometry for those reference stars than CSST, and thus CSST astrometry is mainly limited by centroiding instead of reference frame. 
Hence we can use centroiding uncertainties as a proxy to estimate the astrometric error of the CSST Suvey Camera. 

Here we describe the astrometric model and the criteria for considering a companion to be detected. 
In terms of the small areas of selected deep fields compared with the main sky survey areas, we assume that all of the stars in the synthetic sample are in the CSST's \SI{17500}{deg^{2}} survey areas. 
Supposing $t_{i}$ is the $i$th observation time of one star, $\alpha_{0}, \delta_{0}$ are the R.A. and Decl. of the star, and the CSST barycentric position in the ICRS frame at $t_{i}$ is
\begin{equation}
    \bm{r}(t_{i}) = [x(t_{i}), y(t_{i}), z(t_{i})],
\end{equation}
in the unit of au.
The linear model of the star (i.e. a single star without a companion) at the $i$th observation is expressed as
\begin{equation} \label{linear}
\begin{split}
    \hat{\alpha}_{L,i}^{*} &= \alpha_{0} \cos\delta_{0} + \mu_{\alpha^{*}} (t_{i}-t_{0}) + 
     \varpi f_{\alpha}(t_i)   \\
    \hat{\delta}_{L,i} &= \delta_{0} + \mu_{\delta} (t_{i}-t_{0}) +  \varpi f_{\delta}(t_i),
\end{split}
\end{equation}
where $\mu_{\alpha^{*}}, \mu_{\delta}$ are the proper motion in R.A. and Decl. of the star, and $t_{0}$ is the reference epoch. 
$f_{\alpha}(t_i), f_{\delta}(t_i)$ are the R.A. and Decl. parallax factors expressed as \citep{Wright2009}
\begin{equation}
\begin{split}
   f_{\alpha}(t_i) &= - \bm{r}(t_{i}) \cdot [-\sin\alpha_{0}, \cos\alpha_{0}, 0]^{\mathrm{T}} 
    \\
   f_{\delta}(t_i) &= - \bm{r}(t_{i}) \cdot  [-\cos\alpha_{0}\sin\delta_{0}, -\sin\alpha_{0}\sin\delta_{0}, \cos\delta_{0}]^{\mathrm{T}}
\end{split}
\end{equation}
The astrometry model of the star (i.e. with a companion) at the $i$th observation is expressed as
\begin{equation} \label{Keplerian}
\begin{split}
    \hat{\alpha}_{A,i}^{*} &= \alpha_{0}^{*} + \mu_{\alpha^{*}} (t_{i}-t_{0}) + \varpi f_{\alpha}(t_i)  + ( BX_{i} + GY_{i} ) \\
    \hat{\delta}_{A,i} &= \delta_{0} + \mu_{\delta} (t_{i}-t_{0}) +  \varpi f_{\delta}(t_i) + ( AX_{i} + FY_{i} )
\end{split}
\end{equation}
where $(BX_{i} + GY_{i}),\ (AX_{i} + FY_{i})$ are the reflex motion of the star in R.A. and Decl. induced by the companion.  
$A, B, F, G$ are Thiele-Innes elements \citep{Thiele1883} expressed as 
\begin{equation}
\begin{split}
A &= \alpha_{s} ( \cos \Omega \cos \omega - \sin \Omega \sin \omega \cos i ) \\
B &= \alpha_{s} ( \sin \Omega \cos \omega + \cos \Omega \sin \omega \cos i ) \\
F &= \alpha_{s} ( -\cos \Omega \sin \omega - \sin \Omega \cos \omega \cos i ) \\
G &= \alpha_{s} ( -\sin \Omega \sin \omega + \cos \Omega \cos \omega \cos i ),
\end{split}    
\end{equation}
where $\alpha_{s}$ is the angular semi-major axis of the photocenter expressed in milli-arcseconds (mas), and $i, \Omega, \omega$ are the inclination, the longitude of the ascending node and the argument of pericenter respectively. 
$X_{i}, Y_{i}$ are the elliptical rectangular coordinates in the orbital plane defined as 
\begin{equation}
\begin{split}
    X_{i} &= \cos E_{i} - e \\
    Y_{i} &= \sqrt{1-e^{2}} \sin E_{i},
\end{split}
\end{equation}
and $E_{i}$ is the eccentric anomaly which can be solved from
\begin{equation}
E_{i} - e \sin E_{i} = \frac{2 \pi}{P} t_{i} + M_{0}.
\end{equation}
For every star with a given g band magnitude, we adopt the CSST position uncertainties shown in Figure \ref{fig:comparison}, and the uncertainties for other six bands are scaled by the theoretical reference lower bound of the positional precision $\sigma_{lb}$ for a diffraction-limited image expressed as \citep{Lindegren1978, VanAltena2013, Lindegren2013}
\begin{equation}
    \sigma_{lb}=\frac{1}{\pi} \frac{\lambda}{D} \frac{1}{\mathrm{SNR}}
\end{equation}
where $\lambda$ is the wavelength of the observed photon, $D= $ \SI{2}{m} is the aperture of the Survey Camera on the CSST, and $\mathrm{SNR}$ is the signal-to-noise ratio in the image. 

The synthetic position is generated as follows:
\begin{equation} \label{obs}
\begin{split}
    \alpha_{i}^{*} &= \hat{\alpha}_{A,i}^{*} + \epsilon_{\alpha^{*},i} \\
    \delta_{i} &= \hat{\delta}_{A,i} + \epsilon_{\delta,i},
\end{split}
\end{equation}
where $\epsilon_{\alpha^{*},i}$ and $\epsilon_{\delta,i}$ are white noise with standard deviation of $\sigma_{\alpha^{*}}$ and $\sigma_{\delta}$ respectively.
In the real observation, the field of view is not perfectly aligned with the local equatorial frame, but the CSST data processing team will calibrate the images to make sure that the reduced astrometric data is aligned with the R.A.-Decl. axes\footnote{Refer to the CSST data challenge: https://nadc.china-vo.org/events/CSSTdatachallenge/info/challenge\_11th}.

\subsection{Selection of unresolvable companions}
Unlike stars with planets, it is necessary to estimate whether substellar binaries are resolvable in the Survey Camera before the yield prediction because substellar binaries tend to have similar masses and thus similar luminosity \citep{Fontanive2018, Factor2023}, which makes the system photocenter shift away from the primary. 

Here we take x band as an example, with x representing any one of the seven photometric bands on the Survey Camera. Given a substellar binary system, the angular semi-major axis of the photocentric motion is
\begin{equation}
    a_{\rm{ph},x} = \frac{a_{\rm{rel}}}{d} \frac{m_{2}}{m_{1} + m_{2}} (1 - \frac{m_{1}F_{2,\rm{x}}}{m_{2}F_{1,\rm{x}}}) (1 + \frac{F_{2,\rm{x}}}{F_{1,\rm{x}}})^{-1}
\end{equation}
where $m_{1}$ and $m_{2}$ are the masses of the primary and the secondary respectively, $F_{1,\rm{x}}$ and $F_{2,\rm{x}}$ are the observed fluxes of the primary and the secondary in x band, $a_{\rm{rel}}$ is the binary semi-major axis, and $d$ is the distance of the system (see equation 1 in \cite{Feng2024eq1}).

The diffraction limit in x band is expressed as 
\begin{equation}
    \theta_{\rm{x}} \approx 1.22 \frac{\overline{\lambda_{\rm{x}}}}{D},
\end{equation}
where $\overline{\lambda_{\rm{x}}}$ is the weighted-average wavelength of x band, and $D= $ \SI{2}{m} is the aperture of the Survey Camera on the CSST. 

If $a_{\rm{ph},x}$ is larger than $\theta_{\rm{x}}$, the binary is resolvable in the x band of the Survey Camera; otherwise it is unresolvable. If the angular semi-major axis of the photocentric motion $a_{\rm{ph}}$ is smaller than the diffraction limit $\theta$ in each band of the Survey Camera, the binary system is totally unresolvable and can only be detected by astrometry; otherwise the system is resolvable.

For a resolvable binary system, if one source is brighter than the limiting magnitude while another is fainter than that, the system can be detected by astrometry. 
If both sources in the resolvable system are brighter than the limiting magnitude, the system can be detected by imaging. 
We don't count imaging-detected brown dwarf binaries as astrometric yield because they might be studied by other groups (see Figure \ref{fig:workflow_BDs} and Table \ref{table: resolvable?} for details).

\begin{table*}[!ht]
\begin{center}
\caption{Three scenarios for whether a BD binary is resolvable (x band as a case). \label{table: resolvable?}}
\begin{tabular}{ccccc}
\hline
Scenario& Resolvable& Component brightness& Detection method& Astrometric yield\\
\hline
$a_{\rm{ph},x} < \theta_{\rm{x}}$ & No  & Blended as one source & Astrometry & Yes \\
$a_{\rm{ph},x} \geq \theta_{\rm{x}}$& Yes & One brighter and another fainter than limiting mag & Astrometry & Yes \\
$a_{\rm{ph},x} \geq \theta_{\rm{x}}$& Yes & Both brighter limiting mag & Imaging & No \\
\hline
\end{tabular}
\end{center}
\end{table*}

\subsection{Statistical criteria for signal selection}
We assume no correlation between R.A. and Decl. and calculate the Bayesian Information Criterion (BIC) for the linear model and astrometry model following the method in \cite{Feng2017} as below
\begin{equation}
\begin{split}
\mathrm{BIC_{0}} &= \chi^{2}_{0, \mathrm{min}} + k_{0} \ln (n) = \sum_{i=0}^{N}[\frac{(\alpha_{i}^{*}-\hat{\alpha}_{L,i}^{* \prime})^{2}}{\sigma_{\alpha^{*}}^{2}}+\frac{(\delta_{i}-\hat{\delta}_{L,i}^{\prime})^{2}}{\sigma_{\delta}^{2}}] + k_{0} \ln(n) \\
\mathrm{BIC_{1}} &= \chi^{2}_{1, \mathrm{min}} + k_{1} \ln (n) = \sum_{i=0}^{N}[\frac{(\alpha_{i}^{*}-\hat{\alpha}_{A,i}^{*})^{2}}{\sigma_{\alpha^{*}}^{2}}+\frac{(\delta_{i}-\hat{\delta}_{A,i})^{2}}{\sigma_{\delta}^{2}}] + k_{1} \ln(n),
\end{split}
\end{equation}
where $k_{0}=5,\ k_{1}=12$ are the number of parameters estimated by the model and $n$ is the number of observations. 
The derivations of $\hat{\alpha}_{L,i}^{* \prime}$ and $\hat{\delta}_{L,i}^{\prime}$ are put in the Appendix.
Finally, we calculate the $\ln \mathrm{BF}$ for each set of orbital parameters of a star as
\begin{equation}
\begin{split}
    \ln \mathrm{BF} = \frac{\mathrm{BIC_{0}} - \mathrm{BIC_{1}}}{2}.
\end{split}
\end{equation}
If $\ln \mathrm{BF}$ is larger than 5, this companion signal is considered to be detected \citep{Kass1995}. 
For every host, the detection rate $r_{j,n}$ in each grid is the ratio between the number of recovered companions and the number of injected companions. 
Moreover, in order to avoid overfitting, only sources with more than six observation times ($n \ge 6$) are included to calculate the yield.
In terms of the criteria of $\ln \mathrm{BF} > 5$ and $n \ge 6$, we calculate the total yield of companions for CSST stellar and substellar sources. 
The total yield $Y_1$ of the synthetic sample of CSST stellar sources is expressed as
\begin{equation}
    Y_{1} = \sum_{k=1}^{N_{\mathrm{star}}} \sum_{j,n}^{N_{\mathrm{grid}}} [f_{j,n}(\log(P),m) \Delta \log(P)_{j} \Delta \log(m)_{n}] \ r_{j,n}(\log(P),m | k)
\end{equation}
where $f_{j,n}$ is the occurrence rate in each $P-m$ grid, and $r_{j,n}$ is the detection rate in each grid for the $k$th star. 
Similarly, the total yield $Y_2$ of the synthetic sample of CSST substellar sources is expressed as
\begin{equation}
    Y_{2} = \sum_{k=1}^{N_{\mathrm{BD}}} \sum_{j,n}^{N_{\mathrm{grid}}} [R_{j,n}(\log(\rho), q) \Delta \log(\rho)_{j} \Delta q_{n}] \ s_{j,n}(\log(\rho), q | k)
\end{equation}
where $R_{j,n}$ is the brown dwarf companion frequency in each $\log(\rho)-q$ grid, and $s_{j,n}$ is the detection rate in each grid for the $k$th brown dwarf. 

\section{Results}
\label{sec:results}
By injecting and recovering substellar companions following the procedures introduced in Section \ref{sec:model}, we obtain the total yield of the synthetic sample of CSST stellar and substellar sources by CSST astrometry. 
The predicted companion yields of FGK-dwarfs, M-dwarfs and brown dwarfs are summarized in Table \ref{table: yield summary}. 
Predicted companion yield for FGK-dwarfs detected by CSST astrometry is scarce, nearly zero.
The reason is that nearby FGK-type stars are typically too bright and thus get saturated in CSST detectors. 

\begin{figure*}[!ht]
\centering
\includegraphics[width=0.6 \textwidth]{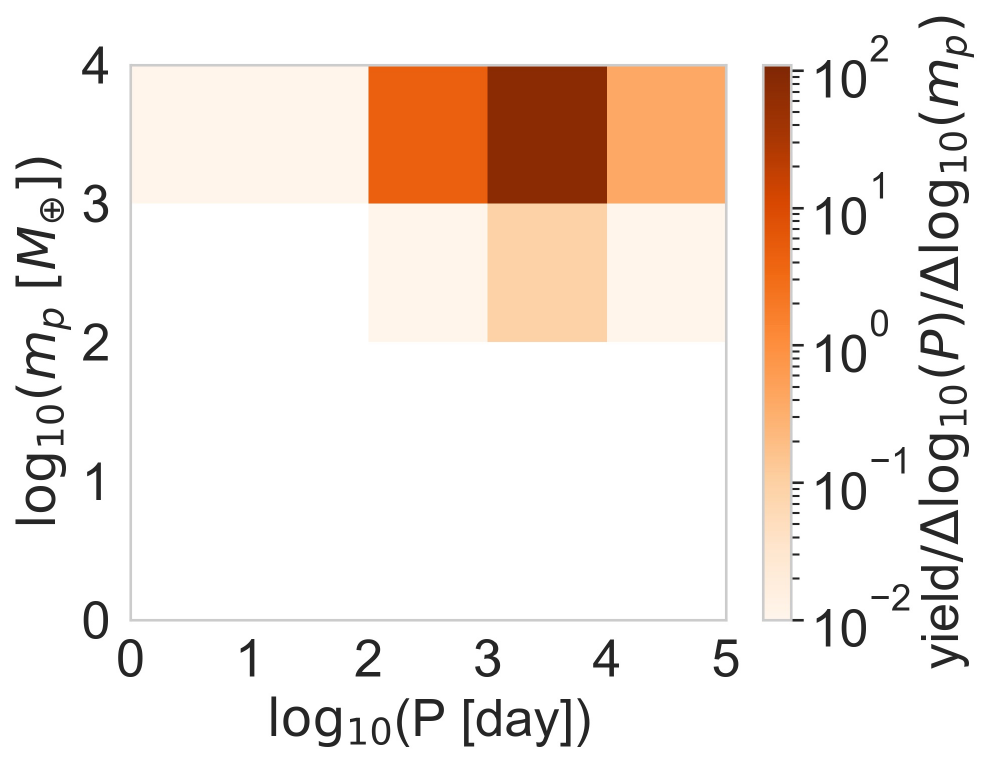}
\caption{Distribution of predicted companion yield in period-companion mass space for M-dwarfs.}
\label{fig:M-dwarfs_2d_yield}
\end{figure*}

\begin{figure*}[!ht]
\centering
\includegraphics[width=0.6 \textwidth]{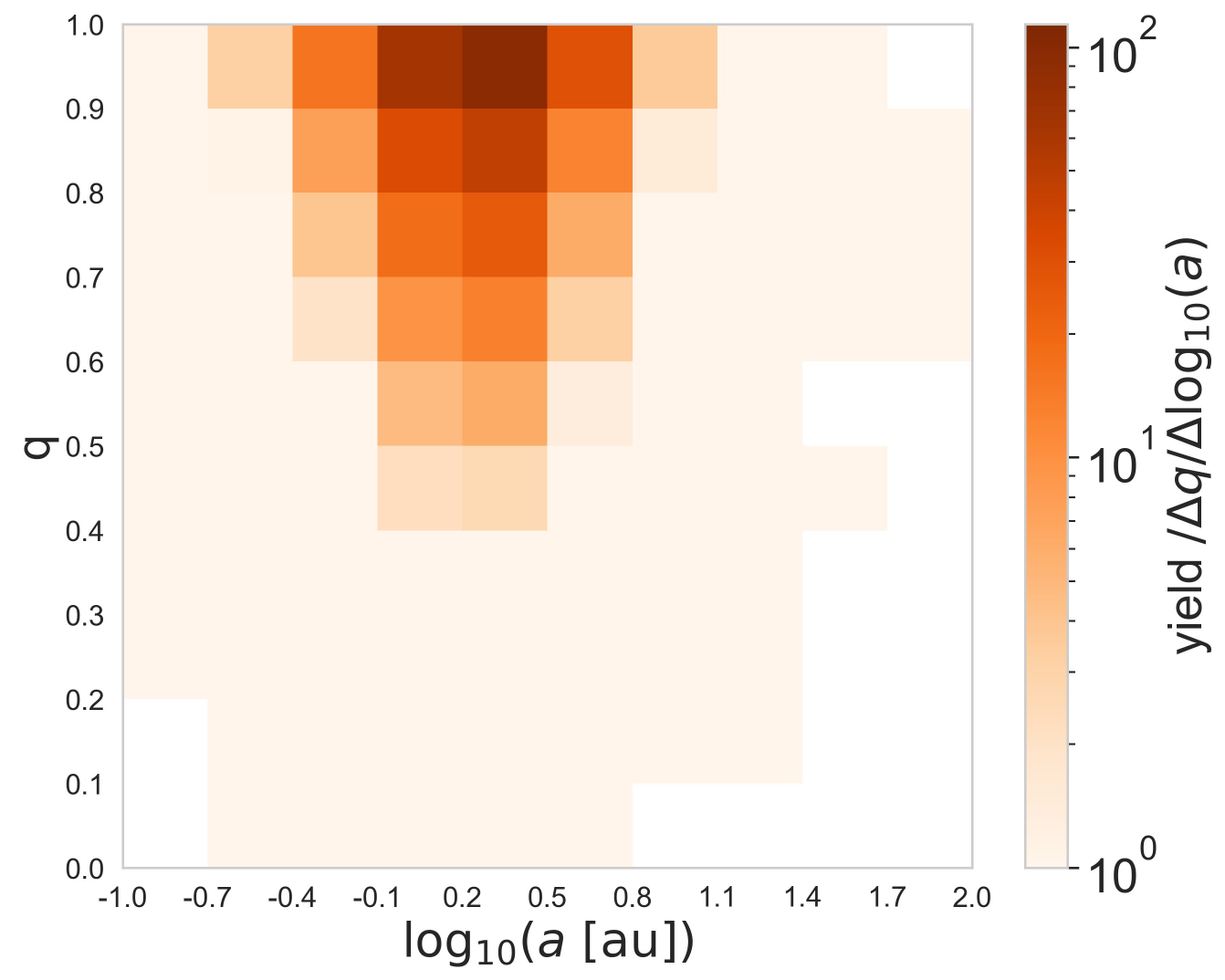}
\caption{Distribution of predicted companion yield in projected separation-mass ratio space for brown dwarfs.}
\label{fig:BDs_2d_yield}
\end{figure*}

\begin{figure*}[!ht]
\centering
\includegraphics[width=1.0 \textwidth]{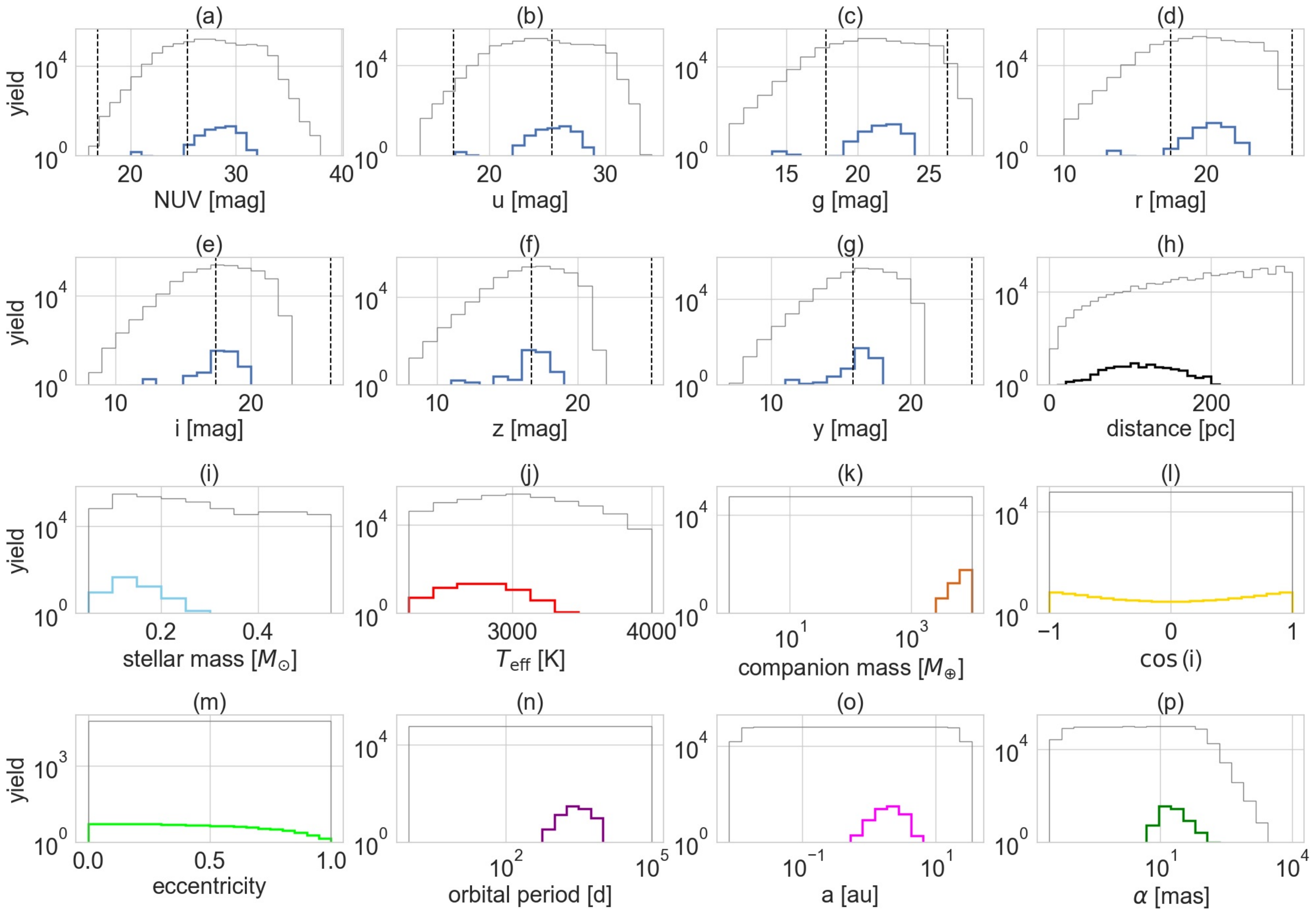}
\caption{Distributions of the M-dwarf yield over parameters. The panels represent parameters as follows: stellar magnitude in NUV, u, g, r, i, z and y band, distance, stellar mass, effective temperature, companion mass, cosine of inclination, eccentricity, orbital period, semi-major axis and angular semi-major axis.
To compare with the companion yield, the gray lines in panels are plotted to show the property distributions of the FGK-dwarf sample or injected companions. 
The range between two dashed vertical lines is the magnitude range that the CSST-SC could observe.}
\label{fig:M-dwarfs_yield}
\end{figure*}

\begin{figure*}[!ht]
\centering
\includegraphics[width=1.0 \textwidth]{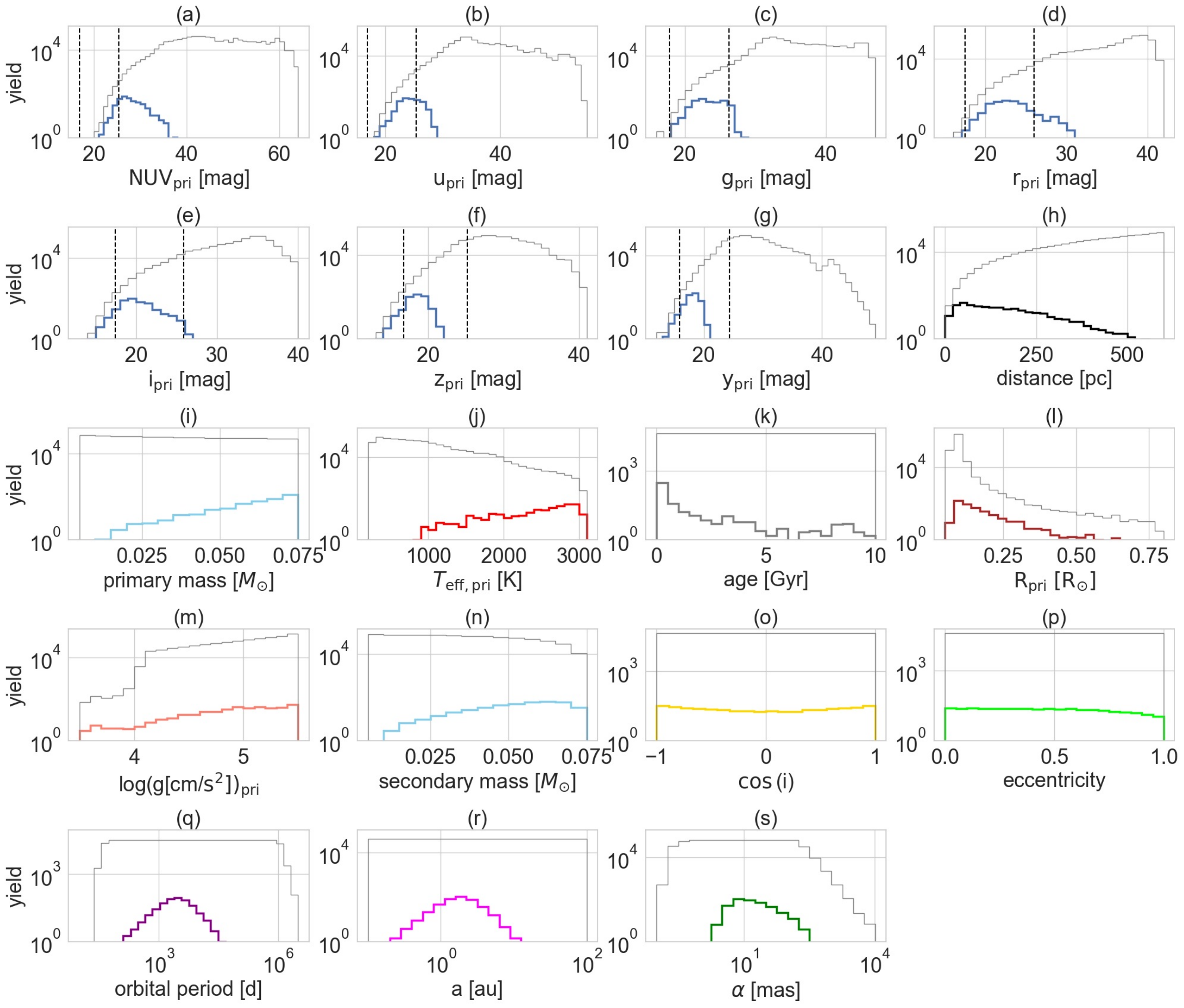}
\caption{Distributions of the BD yield over parameters. The panels represent parameters as follows: primary magnitude in NUV, u, g, r, i, z and y band, distance, primary mass, effective temperature of primaries, system age, primary radius, log10 of the surface gravity for primaries, secondary mass, cosine of inclination, eccentricity, orbital period, semi-major axis and angular semi-major axis. 
To compare with the companion yield, the gray lines in panels are plotted to show the property distributions of the brown dwarf sample or injected companions. 
The range between two dashed vertical lines is the magnitude range that the CSST-SC could observe.}
\label{fig:BDs_yield}
\end{figure*}

The 2D distributions of predicted companion yield in orbital period-companion mass (or projected separation-mass ratio) space are shown in Figure \ref{fig:M-dwarfs_2d_yield} and \ref{fig:BDs_2d_yield} for M-dwarfs and BDs respectively.
The histograms of predicted companion yield detected by CSST astrometry, as a function of various stellar and companion physical quantities, are shown in Figure \ref{fig:M-dwarfs_yield} and \ref{fig:BDs_yield} for M-dwarfs and BDs respectively. 

\begin{table*}[!ht]
\begin{center}
\caption{Yields of substellar companions for three types of hosts. \label{table: yield summary}}
\begin{tabular}{ccccc}
\hline
Host type& Total& Planet companions& BD companions& Occurrence rate model\\
\hline
FGK-dwarfs& $\sim0$ & $\sim0$ & $\sim0$ & \citet{Fernandes2019}\\
M-dwarfs& $83^{+84}_{-62}$ & $6^{+6}_{-4}$ & $77^{+78}_{-58}$ & \citet{Clanton2016}\\
brown dwarfs& $420^{+153}_{-115}$ & $2^{+1}_{-1}$ & $418^{+152}_{-114}$ & \citet{Factor2023}\\
\hline
\end{tabular}
\end{center}
\end{table*}

The number of predicted substellar companions around M-dwarfs that CSST could discover via astrometry in 10 years is $83^{+84}_{-62}$, among which include $77^{+78}_{-58}$ brown dwarfs and $6^{+6}_{-4}$ giant planets. 
The large uncertainties of the companion yield mainly come from 68\% confidence intervals of the occurrence rate in table 3 of \cite{Clanton2016}. 
As in Figure \ref{fig:M-dwarfs_2d_yield}, the majority of those companions are located in the range of $10^{3}-10^{4} \ \mathrm{days}$ and $10^{3}-10^{4} \ M_{\oplus}$ because this region of the parameter space is where CSST is the most sensitive to given its expected astrometric precision and operation time. 
For substellar companions with masses less than $10^{3} \ M_{\oplus}$, the astrometric signatures are too small for the CSST-SC to detect. 
In addition, substellar companions with orbital periods less than $10^{3} \ \mathrm{days}$ and longer than $10^{4} \ \mathrm{days}$ are rarely detected because of the sparse observing cadence (${\sim}200 \ \mathrm{days}$) and limited operation time (${\sim}10 \ \mathrm{yr}$) of CSST respectively.
In the panel (h) of Figure \ref{fig:M-dwarfs_yield}, the M-dwarf companion yield peaks at $\sim{100}$\,pc and decreases to zero beyond 200\,pc.
In the panel (m), the number of detected companions with eccentric orbits is less than the number of companions with circular orbits because the eccentric companions stay at the position around the apocentre for a long time in one orbital period, which might be mistakenly regarded as low eccentricity via astrometry.

The number of predicted companions around brown dwarfs that CSST could discover via astrometry in 10 years is $420^{+153}_{-115}$, among which include $418^{+152}_{-114}$ brown dwarfs and $2 \pm 1$ giant planets. 
The majority of them are secondary brown dwarfs in binary systems. 
The source of uncertainties is from the error propagation of the companion frequency uncertainty $F=0.11^{+0.04}_{-0.03}$ in Equation \ref{eq: BD companion freq}. 
As in Figure \ref{fig:BDs_2d_yield}, the majority of brown dwarf companions are located in the range of $0.4-12.6 \ \rm{au}$ and $q \geq 0.6$ in projected separation-mass ratio space, which indicates that brown dwarf binaries tend to have tightly bound orbits and high mass ratios.
In the panel (h) of Figure \ref{fig:BDs_yield}, the BD companion yield peaks at $\sim{50}$\,pc and decreases to zero at $\sim{500}$\,pc.
In the panel (k), the BD companion yield peaks at the young age ($\leq1$\,Gyr) and gradually decreases over time because young brown dwarfs are relatively bright and thus easier to be detected by astrometry.
In the panel (q), the orbital period of BD binaries ranges from $10^{2}$ to $10^{5}$ days.
In the panel (o), the number of detected companions with the inclination of ${\sim}0 \ \mathrm{deg}$ or ${\sim}180 \ \mathrm{deg}$ is more than the number of companions with the inclination of ${\sim}90 \ \mathrm{deg}$ because the astrometry method is more sensitive to face-on systems than edge-on systems.


\section{Conclusions}
\label{sec:conclusions}
In this work, we investigate the ability of CSST to detect substellar companions via differential astrometry and predict the number of companions detectable with the CSST-SC. 
We predict that CSST could barely discover giant planets and low-mass BDs around FGK-dwarfs, but it could detect $83^{+84}_{-62}$ giant planets and low-mass BDs around M-dwarfs within \SI{300}{pc}, and $420^{+153}_{-115}$ brown dwarf binaries within \SI{600}{pc} by astrometry during its 10-year mission.

CSST astrometry occupies a unique niche among the operating and upcoming space-based survey projects. 
While the final astrometric precision of the CSST-SC does not surpass that of Gaia for objects brighter than 21 mag due to less observing cadence, CSST excels in measuring fainter stars owing to its superior sensitivity. 
Compared with the Nancy Grace Roman Space Telescope \citep{Sanderson2019, Gandhi2023}, the CSST-SC has an equivalent single-exposure astrometric precision. 
However, CSST exceeds Roman in both field of view (\SI{1.1}{deg} versus \SI{0.28}{deg}) and sky survey area (\SI{17500}{deg^{2}} versus \SI{2000}{deg^{2}}), enabling it to observe a significantly broader distribution of target systems. 


It should be noted that CSST probably has 10-20\% time in its 10-year lifespan which will open for proposals. 
A dedicated proposal of observing a smaller sky area with high cadence will facilitate high-precision differential astrometry and enable CSST to produce astrometric precision similar to Gaia, making it more likely detect smaller planets with shorter orbital periods.
Furthermore, it is worthwhile to combine the astrometric data of CSST with Gaia data to extend the observational baseline and detect longer-period substellar companions around nearby stars and brown dwarfs. 

CSST astrometry will provide a precious sample of giant planet and BD companions around nearby M-dwarfs and brown dwarfs, which will further constrain the occurrence rate of long-period massive companions around low-mass stars and BDs, improve our understanding of their formation and evolution channels, and complement the catalogue of objects in the solar neighborhood. 
\clearpage

\section*{Appendix}
For a target star, here we fit the observation data points with a linear trend and re-derive the five parameters $[\alpha^{*}_{0}, \delta_{0}, \mu_{\alpha_{*}}, \mu_{\delta}, \varpi]^{\mathrm{T}}$ of the star in the linear model as 
\begin{equation}
    \bm{\theta} = [\alpha^{* \prime}_{0}, \delta^{\prime}_{0}, \mu^{\prime}_{\alpha^{*}}, \mu^{\prime}_{\delta}, 
    \varpi^{\prime}]^{\mathrm{T}},   \tag{A1} 
\end{equation}
and the vector of the observation data points for the star is
\begin{equation}
    \bm{b} = [\alpha_{1}^{*}, \delta_{1},  ..., \alpha_{n}^{*}, \delta_{n}]^{\mathrm{T}},  \tag{A2}
\end{equation}
where $n$ is the number of times that the target star will be observed by CSST-SC.
The $2n \times 5$ coefficient matrix is expressed as 
\begin{equation}
    \bm{A} = \begin{bmatrix}
        1 & 0 & t_{1}-t_{0} & 0 & f_{\alpha}(t_{1})\\
        0 & 1 & 0 & t_{1}-t_{0} & f_{\delta}(t_{1})\\
        \vdots & \vdots & \vdots & \vdots & \vdots \\
        1 & 0 & t_{n}-t_{0} & 0 & f_{\alpha}(t_{n})\\
        0 & 1 & 0 & t_{n}-t_{0} & f_{\delta}(t_{n})  \tag{A3}
    \end{bmatrix},
\end{equation}
and assuming that the measurement of R.A. and Decl. is independent, the $2n \times 2n$ covariance matrix is expressed as 
\begin{equation}
    \bm{\Sigma} = \begin{bmatrix}
        \sigma_{\alpha^{*},1}^{2} & & & & \\
         & \sigma_{\delta,1}^{2} & & &  \\
         & & \ddots & & \\
         & & & \sigma_{\alpha^{*},n}^{2} & \\
         & & & & \sigma_{\delta,n}^{2}  \tag{A4}
    \end{bmatrix}.
\end{equation}
Thus $\chi^{2}$ is expressed as
\begin{equation}
    \chi^{2} = (\bm{A \theta} - \bm{b})^{\mathrm{T}} \bm{\Sigma^{-1}} (\bm{A \theta} - \bm{b}),  \tag{A5}
\end{equation}
and by differentiating $\chi^{2}$ with respect to $\bm{\theta}$, we derive
\begin{equation}
    \bm{\theta} = (\bm{A}^{\mathrm{T}} \bm{\Sigma}^{-1} \bm{A})^{-1} \bm{A}^{\mathrm{T}} \bm{\Sigma}^{-1} \bm{b} 
    = [\alpha^{* \prime}_{0}, \delta^{\prime}_{0}, \mu^{\prime}_{\alpha^{*}}, \mu^{\prime}_{\delta}, \varpi^{\prime}]^{\mathrm{T}},  \tag{A6}
\end{equation}
so the linear model is re-expressed as
\begin{equation} 
\begin{split}
    \hat{\alpha}_{L,i}^{* \prime} &= \alpha_{0}^{* \prime} + \mu_{\alpha^{*}}^{\prime} (t_{i}-t_{0}) + 
    \varpi^{\prime}f_{\alpha}(t_{i}) \\
    \hat{\delta}_{L,i}^{\prime} &= \delta_{0}^{\prime} + \mu_{\delta}^{\prime} (t_{i}-t_{0}) +
    \varpi^{\prime}f_{\delta}(t_{i}).
\end{split}   \tag{A7}
\end{equation}

\section*{Acknowledgments}
We thank the referees for their insightful comments and valuable suggestions that greatly improved our paper.
We thank Xianmin Meng for providing the CSST filter transmission curves, and Youhua Xu for providing the pointing center time-series data of the CSST Survey Camera.
We also thank Kaiming Cui, Guangyao Xiao, Yicheng Rui, Zhongze Li, and Patrick Tamburo for helpful discussions.
This work is supported by the National Key R\&D Program of China, No. 2024YFC2207700 and 2024YFA1611801, by the National Natural Science Foundation of China (NSFC) under Grant No. 12473066, by the Shanghai Jiao Tong University 2030 Initiative, and by the China-Chile Joint Research Fund (CCJRF No. 2205). 
CCJRF is provided by the Chinese Academy of Sciences South America Center for Astronomy (CASSACA) and established by the National Astronomical Observatories, Chinese Academy of Sciences (NAOC) and Chilean Astronomy Society (SOCHIAS) to support China-Chile collaborations in astronomy. 
S.L. acknowledges support from the National Key R\&D Program of China (Grant No. 2023YFA1607901), the Youth Innovation Promotion Association CAS with Certificate Number 2022259, the Talent Plan of Shanghai Branch, Chinese Academy of Sciences with No. CASSHB-QNPD-2023-016.
Y.C. acknowledges support from the Natural Science Research Project of Anhui Educational Committee No. 2024AH050049, National Natural Science Foundation of China (NSFC) No. 12003001, the China Manned Space Project (Grant No. CMS-CSST-2021-A08), the Anhui Project (Z010118169).
We acknowledge the science research grants from the China Manned Space Project with NO.CMS-CSST-2021-A12, NO.CMS-CSST-2021-B10. 
The computations in this paper were run on the Siyuan-1 cluster supported by the Center for High Performance Computing at Shanghai Jiao Tong University.

\section*{Author Contributions}
Fabo Feng provided the original concept of this work. 
Yifan Xuan led this work, built synthetic samples, constructed occurrence rate models and astrometric models, ran simulations, interpreted results and wrote the manuscript with help mainly from Fabo Feng.
Zhensen Fu, Shilong Liao and Zhaoxiang Qi provided the data of CSST-SC's astrometric precision.
Yang Chen provided the Milky Way mock stellar catalogue for the CSST-SC generated from the \verb|TRILEGAL| code.
All authors reviewed and contributed to this manuscript.

\software{numpy \citep{numpy2020}, 
pandas \citep{pandas2022}, 
scipy \citep{scipy2020},
matplotlib \citep{Hunter2007}, 
astropy \citep{astropy2013, astropy2018, astropy2022}}

\bibliography{sample631}{}
\bibliographystyle{aasjournal}



\end{document}